\documentclass[a4paper,pra,onecolumn,showpacs,superscriptaddress,groupedaddress]{revtex4}
\usepackage{ae}
\usepackage{physics}
\usepackage{amsthm}
\usepackage[colorlinks,citecolor=blue,urlcolor=blue,bookmarks=false,hypertexnames=true]{hyperref}
\usepackage[T1]{fontenc}
\usepackage[ansinew]{inputenc}
\usepackage{amsmath}
\usepackage{amssymb}
\usepackage[caption=false]{subfig}
\usepackage{multirow}
\usepackage{array}
\usepackage[]{graphicx}
\usepackage{wrapfig}
\usepackage{makecell}
\usepackage{color}
\usepackage[colorlinks]{hyperref}
\usepackage{lscape}
\theoremstyle{plain}

\theoremstyle{definition}

\theoremstyle{remark}

\graphicspath{ {./images1/} }

\begin{document}
\title{Structure of Resource Theory of Block Coherence}
\author{Prabir Kumar Dey}
\email{prabirkumardey1794@gmail.com}
\affiliation{Department of Applied Mathematics, University of Calcutta, 92 A.P.C Road, Kolkata 700009, India}
\author{Dipayan Chakraborty}
\email{dipayan.tamluk@gmail.com}
\affiliation{Department of Mathematics, Sukumar Sengupta Mahavidyalaya, Keshpur, Paschim Medinipur, 721150, India}
\author{Priyabrata Char}
\email{mathpriyabrata@gmail.com}
\affiliation{Department of Applied Mathematics, University of Calcutta, 92 A.P.C Road, Kolkata 700009, India}
\author{Indrani Chattopadhyay}
\email{icappmath@caluniv.ac.in}
\affiliation{Department of Applied Mathematics, University of Calcutta, 92 A.P.C Road, Kolkata 700009, India}
\author{Amit Bhar}
\email{bharamit79@gmail.com}
\affiliation{Department of Mathematics, Jogesh Chandra Chaudhuri College, 30, Prince Anwar Shah Road, Kolkata 700033, India}
\author{Debasis Sarkar}
\email{dsarkar1x@gmail.com, dsappmath@caluniv.ac.in}
\affiliation{Department of Applied Mathematics, University of Calcutta, 92 A.P.C Road, Kolkata 700009, India}

\begin{abstract}
Emerging from the superposition principle, the resource theory of coherence plays a crucial role in many information-processing tasks. Recently, a generalization to this resource theory was investigated with respect to arbitrary positive operator valued measurement (POVM) based on Naimark's dilation theorem. Here, we introduce the notion of Block Incoherent Operations (BIO), Strictly Block Incoherent Operations (SBIO) and  Physically Block Incoherent Operations (PBIO) and provide an analytical expression for Kraus operators of these operations to have a better understanding of the resource theory of block coherence which in turn gives a more clear picture of POVM based resource theory of coherence. A dilation theorem corresponding to SBIO has been introduced to enlighten the proper physical interpretation of this operation. These free operations will be helpful in finding out the conditions of state transformations and could be implemented in various protocols. For a transparent view of this resource theory, we have successfully introduced the concept of state transformation under SBIO.
\end{abstract}
\date{\today}
\pacs{ 03.67.Mn.; 03.65.Ud.}
\maketitle
{Keywords: Coherence; Resource Theory; Generalized Measurement; Block Coherence.}
\section{Introduction}
Quantum Resource theories \cite{1,2} provide a framework to perform certain types of tasks that are otherwise not possible by the laws of classical physics. This provides a boundary between the classical world and the quantum world. Resource theories in quantum information assist us in observing quantum resources in numerous physical phenomena and help us to develop various important protocols. The basic construction of resource theory involves free states and free operations \cite{3,4,5,6}. Free operations represent physical transformations that can be implemented without the utilization of resources, while free states can be constructed without any additional cost. States that can not be prepared from free states under free operations are called resource states, and these states are widely used in respective resource theories to perform various information processing tasks \cite{7,8,9}. In entanglement theory, free states are separable states, and free operations are Local operations along with classical communications (LOCC).  \\
The resource theory of coherence has been identified as an important field of research in recent times \cite{10}. Resource theory of coherence mainly originated from the superposition principle of quantum states. The basic structure of this resource theory consists of incoherent states as free states and incoherent operations as free operations. As the superposition principle is basis-dependent, we have to choose a fixed basis to develop the resource theory of coherence.  Quantum coherence has been captured as a resource behind various physical phenomena. Its applications include biological systems \cite{cobio 1, cobio 2, cobio 3, cobio 4}, quantum thermodynamics \cite{cothermo 1,cothermo 2,cothermo 3,cothermo 4}, quantum metrology \cite{cometrology 1, cometrology 2, cometrology 3, cometrology 4} etc. Protocols like distillation \cite{11}, dilution \cite{11}, assisted distillation \cite{coherence assisted distillation 2, coherence assisted distillation 1}, and incoherent quantum state merging \cite{coherence state merging}, quantum phase discrimination \cite{phase discrimination1, phase discrimination2}, sub-channel discrimination \cite{sub-channel discrimination} and various catalysis procedure involving coherence \cite{catalytic coherence 1, catalytic coherence 2, catalytic coherence 3, char catalysis 1, char catalysis 2} etc. have also been studied within coherence resource theory.\\
In quantum theory, measurement plays an important role. Recently, the resource theory of coherence based on Positive Operator Valued Measurement (POVM) has been studied in \cite{15}. This approach provides us with a framework to understand coherence more fundamentally, as POVMs are the most general kind of quantum measurement. The POVM framework is based on the Naimark dilation theorem, which states that every POVM can be extended to higher-rank projective measurements by embedding it into a larger Hilbert space. Higher rank projective measurements correspond to the resource theory of block coherence in which the block diagonal states are considered free states. In the case of projective measurements of rank one, this approach coincides with the standard resource theory of coherence. Now, in order to develop any resource theory, we need to have a better understanding of free operations. Different incoherent operations were studied in literature with respect to the standard resource theory of coherence \cite{16,17, 18}. In this work, we have concentrated on various incoherent operations with respect to the resource theory of block coherence. In this regard, we have introduced a state transformation protocol under a free operation of block coherence, in particular, Strictly Block Incoherent Operation (SBIO). As SBIO forms a subclass of Block Incoherent Operation (BIO), our result will also be significant under BIO too. Our result is able to detect the output states under the action of SBIO from an input pure state under certain considerations. This will certainly make the resource theory of block coherence more physically significant. We also compare standard coherence theory and block coherence theory regarding state transformation. \\
This paper is organized as follows. In section 2, we briefly review the standard resource theory of coherence, Block coherence and POVM-based resource theory of coherence. In sections 3, 4 and 5, we introduce three different block incoherent operations viz. \textbf{Block Incoherent Operations (BIO)}, \textbf{Strictly Block Incoherent Operations (SBIO)}, and \textbf{Physically Block Incoherent Operations (PBIO)}, respectively. We have also given an analytical formulation of their Kraus operators. We introduce concepts of state transformation under SBIO In section 6. In section 7, we provide some physical evidence and applications to enrich the structure of coherence theory. Finally, section 8 ends with the conclusion. \\

\section{Preliminaries}

One of the fundamental differences between entanglement theory and coherence theory is the uniqueness of free operation. In coherence theory, Incoherent Operations (IO), Strictly Incoherent Operations (SIO), Physically Incoherent Operations (PIO), and Maximally Incoherent Operations (MIO) are treated as free operations that are generated from different approaches and motivations fulfilling the minimum requirements for being free operations.

Due to the dependence of basis, we first fix some ordered basis $\{\ket{i}\}_{i=0}^{d-1}$ for a $d$ dimensional system.

Any diagonal state with respect to this basis set is called an incoherent state which has the form $$\rho=\sum_{i=0}^{d-1} p_i\ket{i}\bra{i}$$ with probabilities $p_i$. Here we consider $\mathcal{I}$  as the set of all incoherent states.

Maximally Incoherent Operations (MIO) \cite{19} are considered the largest class of free operations. A Completely Positive Trace Preserving(CPTP) map $\Lambda$ is an MIO if for each $\rho\in \mathcal{I}$ we have $\Lambda[\rho]\in \mathcal{I}$.
For Incoherent Operations (IO) \cite{10} and Strictly Incoherent Operations (SIO) \cite{20}, the CPTP map has proper Kraus representation, but these operations have originated from different physical significance.  Physically Incoherent Operations (PIO) \cite{18} is considered the most significant and relevant free operation of coherence theory.\\
Let, $\Lambda$ has Kraus operator representation $\{K_n\}_n $. Then $\Lambda$ is called an Incoherent Operation (IO) \cite{10} if for each n,  we have $\frac{K_n\rho K_n^\dagger}{Tr[K_n\rho K_n^\dagger]}\in \mathcal{I}$ whenever $\rho\in \mathcal{I}$, i.e., coherence is not generated even probabilistically. This map $\Lambda$ is an Strictly Incoherent Operation (SIO) \cite{20}, if further $K_n\delta(\rho) K_n^\dagger=\delta(K_n\rho K_n^\dagger)$, where we consider $\delta$ as completely dephasing map, i.e., for any state $\rho$,
$$\delta [\rho]=\sum_{i=0}^{d-1}\bra{i}\rho \ket{i} \ket{i}\bra{i}.$$
Every physical operation consists of four fundamental operations, viz., adding some ancillary system, unitary evolution of the joint system, projective measurement and tracing out a subsystem. Hence, Physically Incoherent Operations (PIO) \cite{18} are defined in four steps: (1) adding an incoherent ancillary system B to the original system A, (2) a joint incoherent unitary $U_{AB}$ applied on the total system, (3) An incoherent projective measurement applied on the ancillary system B(generally rank 1 PVM),(4) A classical processing channel applied to the measurement outcomes .\\ 
Clearly, we have PIO $\subseteq$ SIO $\subseteq$ IO $\subseteq$ MIO  \cite{18}.
 
Based on the most general measurement (POVM), a generalization of the resource theory of coherence has been developed recently \cite{15}. They first defined block incoherent states, block coherence measures, and maximally block incoherent operations, and thereafter, they described all of these with respect to POVM using the Naimark extension.
According to Naimark's dilation theorem, every POVM can be extended to a projective measurement $\textbf{P}=\{P_i\}$ by embedding in a larger Hilbert space. The process of $\textit{canonical Naimark extension}$ can be performed by the following steps: (1) attaching ancilla ( generally fixed state $\ket{1}\bra{1})$ to the system state $\rho$ via tensor product, (2) a unitary evolution $V$ of the global system, (3) a Von Neumann projective measurement on the ancillary system.
The probability of obtaining ${i}$th outcome of the projection valued measurement (PVM) on ancilla is $Tr[(\textbf{1}\otimes \ket{i}\bra{i})\rho']$ where $\rho'=V(\rho\otimes \ket{1}\bra{1})V^\dagger$. This probability will be same as for the POVM if $$Tr[E_i\rho]=Tr[(\textbf{1}\otimes \ket{i}\bra{i})\rho']=Tr[P_i(\rho\otimes \ket{1}\bra{1})] $$
$P_i$ being PVM of rank $d$ defined as $P_i=V^\dagger(\textbf{1}\otimes \ket{i}\bra{i})V$. This technique provides a simpler method to implement POVM in any experiment.\\

So, a complete picture of the resource theory of block coherence is very much necessary to deal with the resource theory of coherence based on POVM measurements. \\
 Block dephasing maps are denoted by $\Delta$ and defined as $$\Delta[\rho]=\sum_i P_i \rho P_i$$where each orthogonal projectors $P_i$ can be of any rank within the dimension of embedded Hilbert space. States that are of the form  $\sum_i P_i \rho P_i$ are called block incoherent states \cite{15}, and we denote the set of all block incoherent states by $I_B$. If we consider a full-rank projector i.e., $P=\ket{0}\bra{0}+\ket{1}\bra{1}+\cdots+\ket{d-1}\bra{d-1}=\textbf{1}$, then any states become block incoherent [$\rho=P\rho P$], as there will be only one block in the diagonal of the density matrix. If all $P_i$ are of rank 1, then it coincides with the definition of incoherent states in the standard resource theory of coherence. \\

Maximally Block Incoherent Operations (MBIO) are defined as the largest class of free operations that can not create block coherence \cite{15}. A map $\Lambda_{MBIO}$ lie under this class if 
$$ \Lambda_{MBIO} \left[ I_B \right] \subseteq I_B$$
POVM Maximally Incoherent Operations (POVM-MIO) are free operations that can not increase POVM-based coherence \cite{15}. POVM Maximally Incoherent Operation is denoted by $\Lambda^{\text{POVM}}_{\text{MIO}}$ acts on $S$ and is defined as $\Lambda^{\text{POVM}}_{\text{MIO}}= \mathcal{E}^{-1} \circ \Lambda' \circ \mathcal{E}$ \cite{15}, where $\mathcal{E}$ performs embedding into larger Hilbert space $S'$ by $\mathcal{E}(\rho)= \rho \otimes \ket{1} \bra{1}$ \cite{15}. $\Lambda'$ is a CPTP map which is Maximally Block Incoherent Operation (MBIO) acting on larger space $S'$ and it is also subspace preserving \cite{15} i.e. $\Lambda'\left[S_\mathcal{E}\right]\subseteq S_\mathcal{E}$ for the space of embedded states $ S_\mathcal{E}=\mathcal{E}[S] \subseteq S^{'}$. Finally, $\mathcal{E}^{-1}$ returns back the total system to the original lower dimensional Hilbert space $S$ by tracing out the ancillary system \cite{15}.\\ 

Similar to POVM-MIO, for the study of POVM Incoherent Operations (POVM-IO), POVM Strictly Incoherent Operations (POVM-SIO), POVM Physically Incoherent Operations (POVM-PIO), we need to apply BIO, SBIO, PBIO on the larger space $S'$ respectively. The operation $ \Lambda_{\text{X}}^{\text{POVM}}$ acting on $S$ is defined as $\Lambda_{\text{X}}^{\text{POVM}}=\mathcal{E}^{-1} \circ \Lambda' \circ \mathcal{E}$ where $\text{X}$ is IO/SIO/PIO. $\mathcal{E}$ performs embedding into larger Hilbert space $S'$ by $\mathcal{E}(\rho)= \rho \otimes \ket{1} \bra{1}$ and $\Lambda'$ is a CPTP map which is BIO/SBIO/PBIO acting on larger space $S'$ and also subspace preserving i.e. $\Lambda'\left[S_\mathcal{E}\right]\subseteq S_\mathcal{E}$ for the space of embedded states $ S_\mathcal{E}=\mathcal{E}[S] \subseteq S^{'}$. Finally, $\mathcal{E}^{-1}$ returns back the total system to the original lower dimensional Hilbert space $S$ by tracing out the ancillary system.\\

Now, we  the notion of Block Incoherent Unitary Operation. This is the generalization of Unitary Operation in block formalism. Unitary matrices are always invertible. So, we need to choose blocks properly for Block Incoherent Unitary Operations.\\
We denote the set of all Block Incoherent Unitary operations as $U_{BI}=\sum_k P_{\pi(k)}\;C\; P_k$, where $C$ is any complex matrix. Now $U_{BI}^{\dagger} U_{BI}=I$ gives
\begin{align*}
\sum_{k,k'}    P_{k'} \; C^* \; P_{\pi(k')}\; P_{\pi(k)}\;C\; P_k =I \\
\implies \sum_k  \left( P_{k} \; C^* \; P_{\pi(k)} \right) \left( P_{\pi(k)}\;C\; P_k \right)=I\\
\implies \sum_k  \left( P_{\pi(k)}\;C\; P_k \right)^\dagger \left( P_{\pi(k)}\;C\; P_k \right)=I
\end{align*}

This ensures for each $k$, $ P_{\pi(k)}C P_k $ is a unitary operator.

\section{Block Incoherent Operation}
A CPTP map $\mathcal{T}$ on $S$ (where $S$ is the set of all quantum states of the system Hilbert space $H$) is said to denote Block Incoherent Operation (BIO) if its representation involves a set of Kraus operators \{$K_n$\} such that for all $n$ and $\rho \in \ I_B$, we have $ \frac{K_n\rho K_{n}^{\dagger}}{\textrm{Tr}(K_n\rho K_{n}^{\dagger})} \in I_B$, i.e. even probabilistically block coherence can not be generated from a block incoherent state under this operation. This operation can be redefined as follows:

A CPTP map $\mathcal{T}$ is a Block Incoherent Operation if it can be represented by Kraus operators $\{K_n\}$ such that $$\Delta(K_n \rho K_{n}^{\dagger})=K_n \rho K_{n}^{\dagger}$$  $\forall \rho \in I_B,\;\forall n$ where $\Delta$ is the block dephasing map. From this representation, it is easy to see that if each diagonal block of the state $\rho \in I_B$ is considered separately with zero entries elsewhere, i.e. $B_l= P_l \rho P_l$, then we have 
\begin{equation}
\Delta(K_n B_l K_{n}^{\dagger})=K_n B_l K_{n}^{\dagger}
\end{equation}
Clearly, the choice of projection operators $P_l$ should be the same as the operators used to construct block dephasing map $\Delta$. In this connection, we have the following lemma in the standard coherence theory.\\
$\textbf{\text{Lemma 1:}}$ A CPTP map $\Lambda$ represents IO if and only if it can be represented by Kraus operators $\{K_n\}$ of the form $K_n=\sum _n c_n \ket{f(n)} \bra{n}$ Where $\ket{f(n)}$ is many to one function from basis set onto itself \cite{18}.\\

Next, we prove the following theorem.\\

\vspace*{5pt}
\noindent
{\bf Theorem~1:} 
Let $\mathcal{T}$ be a Completely Positive Trace Preserving (CPTP) map on S. $\mathcal{T}$ denotes Block Incoherent Operation if and only if it can be represented by Kraus operators $\{K_j\}$ of the form $K_j=\sum _{i} P_{f(i)} C_j P_i$, where $f$ is some index function and $C_j$ is a complex matrix.

\vspace*{12pt}
\noindent
{\bf Proof:}
(Partition of columns in $K_n$ can be constructed as follows. First, we need to express $K_n$ in block formalism. For this, we choose each diagonal block of $K_n$ according to $\pi_l= im P_l$, where $P_l$ are projectors defined earlier in the expression of BIO. After choosing diagonal blocks, off-diagonal blocks can be formed automatically. In the block formalism of $K_n$, the blocks appearing vertically in a sequence form one column partition.)\\   
The proof of sufficient part is trivial. If a CPTP map has Kraus operator representation $\{K_n\}$ such that each Kraus operator  has at most one non-zero block in each column partition, then clearly for all $n$ and $\rho \in \ I_B$, we have $ \frac{K_n\rho K_{n}^{\dagger}}{\textrm{Tr}(K_n\rho K_{n}^{\dagger})} \in I_B$.\\
Now, we prove the necessary part. Given $\mathcal{T}$ is Block Incoherent Operation. Suppose each Kraus operators in the representation of $\mathcal{T}$ has the form  $K_n = \sum_{i,j}c_{ij}^n\ket{i}\bra{j} $. We consider $B_l=\sum_{x,y}d_{xy}^{l}\ket{x}\bra{y}$. \\ \\
Then, from (1), we get,
\begin{multline*}
    \hspace{0.4in}\Delta \left[ \sum_{i,j,x,y,i^{'},j^{'}}\left(c_{ij}^n\ket{i}\bra{j}\right) \; \left(d_{xy}^{l}\ket{x}\bra{y}\right)\; \left(c_{i^{'} j^{'}}^{n^*}\ket{j'}\bra{i'} \right) \right]
=\sum_{i,j,x,y,i^{'},j^{'}} \left(c_{ij}^n\ket{i}\bra{j}\right)\;\left(d_{xy}^{l}\ket{x}\bra{y}\right)\;\left(c_{i^{'} j^{'}}^{n^*}\ket{j'}\bra{i'}\right)\\
\implies \Delta\left(\sum_{\substack{i,x,y,i^{'}}} c_{ix}^n d_{xy}^{l} c_{i^{'}y}^{n^*}\ket{i}\bra{i'}\right)=\sum_{\substack{i,x,y,i^{'}}} c_{ix}^n d_{xy}^{l} c_{i^{'}y}^{n^*}\ket{i}\bra{i'}\hspace{1.7in} 
\end{multline*}

So, the above equation must imply off-diagonal blocks
\begin{equation}
\sum_{x,y} c_{ix}^n d_{xy}^{l} c_{i^{'}y}^{n^*}=0
\end{equation}
Here the choice of off-diagonal blocks depends on chosen block dephasing map $\Delta$, i.e., for a suitable choice of $i$ and $i'$.\\ \\
First, for any state $\sigma$ of dimension $d$, we choose $\Delta$ such that $\Delta\left[ \sigma \right]=\sum_{l=0}^1 P_l \sigma P_l$, where $P_0$ and $P_1$ are projectors of rank $d_1$ and $(d-d_1)$ respectively. Clearly $\sum_{l=0}^1 P_l \sigma P_l$ is block incoherent state and it has two diagonal blocks $B_l=P_l \sigma P_l,\;l=\;0,1$. This diagonal blocks give values of $d_{xy}^{l},\;l=\;0,1$. After choosing this block dephasing map, (2) gives two sets of equations given below

\begin{equation}
\sum_{x,y} c_{ix}^n d_{xy}^{l}c^{n^*}_{i^{'}y}=0
\end{equation}
for $i=0$ to $(d_{1}-1)$ and $i^{'}=d_1$ to $(d-1)$. 
\begin{equation}
\sum_{x,y} c_{ix}^n d_{xy}^{l}c^{n^*}_{i^{'}y}=0
\end{equation}
 for $i=d_1$ to $(d-1)$ and for $i^{'}=0$ to $(d_1-1)$. \\ \\
 If we choose $\sigma=\frac{\ket{x}+\ket{y}}{\sqrt{2}}$, where $x$ ranges from $0$ to $(d-1)$ and $y$ ranges from $0$ to $(d-1)$, then values of $d_{xy}^l,\;l=\;0,1$ can be obtained. We substitute these values in a set of equations (3) and (4) properly. After substituting values of $d_{xy}^0$ in a set of equations (3), we can show that if we choose any one element of a diagonal block of order $d_1 \times d_1$ of $K_n$ to be non-zero, then corresponding off-diagonal block below this diagonal block becomes zero and vice versa. Similarly, when we substitute values of $d_{xy}^{1}$ in the set of equations (3), we can easily show that if we choose any one element of the diagonal block of order $(d-d_1) \times (d-d_1)$ of $K_n$ to be non-zero then corresponding off-diagonal block above this diagonal block becomes zero and vice versa.
The same result can be obtained if we substitute values of $d_{xy}^{l}, l=0,1$ in a set of equations (4). \\
Now, one or both the diagonal blocks of order $d_1 \times d_1$ and $(d-d_1) \times (d-d_1)$ of $K_n$ can be further decomposed into fewer dimensional blocks and the same procedure can be followed for each block through the proper implementation of the result just discussed above. The addition of new fewer dimensional blocks clearly depends on the existence of non-zero elements in the previous higher dimensional blocks, i.e., we can repeat the process if we find any new block for which there is no non-zero element in the column partition of the corresponding block. We have the main result by continuing this process for a finite number of times.\\
 We refer to Appendix A for elaborate proof of the five-dimensional case. $\square$\\

\section{Strictly Block Incoherent Operation}
Strictly Block Incoherent Operation (SBIO) are constructed by CPTP maps $\mathcal{T}$ on S having a set of Kraus operators \{$K_n$\} such that for all n and $\rho$, $$ \Delta (K_n \rho K_{n}^{\dagger}) = K_n \Delta (\rho) K_{n}^{\dagger}$$
This can be redefined as follows:\\
Let $\mathcal{T}$ be a CPTP map. Then $\mathcal{T}$ is said to be a Strictly Block Incoherent Operation if Kraus operators representing $\mathcal{T}$ satisfies 
    
\begin{equation}
\Delta(K_n B_l K_{n}^{\dagger})=K_n B_l K_{n}^{\dagger}\;\;\;\;\forall B_l,\;\forall n \;\;\;
\text{ and }
\;\;\; \Delta (K_n B_{kl}^{'} K_{n}^{\dagger})=0 \; \; \; \forall B_{kl}^{'} , \; \forall n 
\end{equation}
where $B_{kl}^{'}$ denotes each off-diagonal block of any state $\rho$ and this off-diagonal block can be formed by $B_{kl}^{'}=P_k \rho P_l$ with $k \neq l$. 
Before establishing our result, we are stating an important lemma on this matter.\\
$\textbf{\textit{Lemma 2:}}$ A CPTP map $\Lambda$ is SIO if and only if it can be represented by Kraus operators $\{K_j\}$ of the form $K_n=\sum_{n} c_n \ket{\pi(n)} \bra {n}$  where $\ket{\pi(n)}$ is permutation from the basis set onto itself \cite{18}.\\
The following theorem gives the analytic structure of Kraus operators representing Strictly Block Incoherent Operations.

\vspace*{12pt}
\noindent
{\bf Theorem~2:} 
: Let $\mathcal{T}$ be a Completely Positive Trace Preserving(CPTP) map on S. $\mathcal{T}$ denotes Strictly Block Incoherent Operation if and only if it can be represented by Kraus operators $\{K_j\}$ of the form $K_j=\sum _{i} P_{\pi(i)} C_j P_i$, where $\pi$ is an index permutation and $C_j$ is complex matrix.

\vspace*{12pt}
\noindent
{\bf Proof:} (The partition of rows and columns will depend on the block dephasing map. The idea of row partitions will be similar to the column partitions defined earlier in the case of BIO.)\\
The sufficient part of the proof follows from the definition of SBIO. If every Kraus operator of a CPTP map has at most one non-zero block in each column and row partition, then clearly it satisfies all the conditions of equation (5) and hence is a SBIO.\\
We now prove the converse part. From the definition, any SBIO is also a BIO. So, by the previous theorem, the Kraus operator representing SBIO must have at most one non-zero block in each column partition. We only need to show the remaining results.\\
Here, we consider, 
$$ K_n= \sum_{i,j} c_{ij}^n \ket{i} \bra{j}  \text{ and } B_{kl}^{'}= \sum_{x,y} f_{x,y}^{kl} \ket{x} \bra{y} \text{ with } k\neq l.$$
For any state $\sigma$ of dimension $d$, we first choose $\Delta$ as $\Delta\left[ \sigma \right]=\sum_{l=0}^1 P_l \sigma P_l$, where $P_0$ and $P_1$ are projectors of rank $d_1$ and $(d-d_1)$ respectively.  \\
Now, from (5), we have 
\begin{equation}
\sum_{x,y} c_{ix}^n f_{xy}^{kl} c_{i'y}^{n^*}=0
\end{equation} for $i=0$ to $d_1-1$, $i'=0$ to $d_1-1$ and 
\begin{equation}
\sum_{x,y} c_{ix}^n f_{xy}^{kl} c_{i'y}^{n^*}=0
\end{equation} for $i=d_1$ to $d-1$, $i'=d_1$ to $d-1$. \\ \\
Here we choose $\sigma=\frac{\ket{x}+\ket{y}}{\sqrt{2}}$, where $x$ ranges from  $0$ to $d_1-1$ and $y$ ranges from $d_1$ to $d-1$, then the off-diagonal blocks $B_{01}^{'}$ and $B_{10}^{'}$ can be formed by $P_0 \sigma P_1$ and $P_1 \sigma P_0$ and subsequently values of $f_{xy}^{01}$ and $f_{xy}^{10}$ can be obtained and we substitute these values in set of equations (6) and (7) properly.
After substituting values of $f_{xy}^{01}$ and $f_{xy}^{10}$ in set of equations (6), we can show that if we choose any one element of diagonal block of order $d_1 \times d_1$ of $K_{n}$ to be non-zero then corresponding off-diagonal block of order $d_1 \times (d-d_1)$ is completely zero and vice versa. Similarly, by substituting values of $f_{xy}^{10}$ and $f_{xy}^{01}$ in set of equations (7), we can show that if we choose any one element of diagonal block of order $(d-d_1) \times (d-d_1)$ of $K_n$ to be non-zero then corresponding off-diagonal block of order $(d-d_1) \times d_1$ becomes completely zero and vice versa. \\
Proceeding like the case of BIO, one or both the diagonal blocks of order $d_1 \times d_1$ and $(d-d_1) \times (d-d_1)$ of $K_n$ can further be decomposed into fewer dimensional blocks and the same procedure can be followed for each block through the proper implementation of the result just discussed above. Decomposition into new fewer dimensional blocks clearly depends on the existence of non-zero elements in the previous higher dimensional blocks, i.e., we can repeat the process if we find any new block for which there is no non-zero element in both the column and row partitions of the corresponding block. We get the desired result by continuing this process for a finite number of times. We refer to Appendix B for elaborate proof of the dimensional case. $\square$\\ 
 Now, we introduce a dilation protocol that can always be constructed in terms of Kraus operators representing SBIO, and this will certainly lead to the proper physical justification of SBIO. 

\vspace*{12pt}
\noindent
{\bf Theorem~3:} 
An operation on a system S is SBIO if it can be constructed from the following elementary processes using an ancilla $\alpha$
\begin{enumerate}
    \item Unitary operations on $\alpha$ controlled by higher rank projectors applied on S: $U=\sum_k P_k\otimes U_k.$ 
    \item Measurements on $\alpha$ in any basis.
    \item Block incoherent unitary operations on S, with respect to the measurement outcome: $V_{\mu}=\sum_i P_{\pi_{\mu}(i)}C^{\mu}P_i.$ 
\end{enumerate}
\vspace*{12pt}
\noindent
{\bf Proof:} We start with some system density matrix $\rho_S$ and add an ancilla $\alpha$ which is initially in some state $\ket{0}\bra{0}_\alpha$. Now, for the evolution, we take the joint unitary as mentioned in the first step. Thus we have,                                            $$U(\rho_S\otimes\ket{0}\bra{0}_\alpha)U^{\dagger}$$
$$=\left(\sum_k P_k\otimes U_k\right) \left(\rho_S\otimes\ket{0}\bra{0}_\alpha\right) \left(\sum_{k'} P_{k'}\otimes U_{k'}^\dagger\right)$$                                  
$$=\sum_{k,k'} P_k \rho_S P_{k'}\otimes U_k \left(\ket{0}\bra{0}_\alpha \right) U_{k'}$$
To get a single outcome $\mu$, we project out a state $\ket{\phi_{\mu}}$ on the ancilla. This leads to\\      
$\bra{\phi_{\mu}}_\alpha\left(\sum_{k,k'} P_k \rho_S P_{k'}\otimes U_k \left(\ket{0}\bra{0}_\alpha \right) U_{k'}\right)\ket{\phi_{\mu}}_\alpha=\sum_{k,k'} P_k \rho_S P_{k'}\bra{\phi_{\mu}}U_k\ket{0}(\bra{\phi_{\mu}}U_{k'}\ket{0})^*.$
Finally, we apply block incoherent unitary $V_{\mu}=\sum_i P_{\pi_{\mu}(i)}C^{\mu}P_i$ on the system S and get,
\begin{equation*}
V_{\mu}\left(\sum_{k,k'} P_k \rho_S P_{k'}\bra{\phi_{\mu}}U_k\ket{0}(\bra{\phi_{\mu}}U_{k'}\ket{0})^*\right) V_{\mu}^\dagger
=K_{\mu} \rho_S K_{\mu}^{\dagger}  
\end{equation*}
where $K_{\mu}=\sum_k \bra{\phi_{\mu}}U_k\ket{0} P_{\pi_{\mu}(k)}C^{\mu}P_k$ clearly satisfies all the conditions for being a Kraus operator of SBIO. All the Kraus operators for this SBIO can be obtained by selecting each of the basis states $\ket{\phi_{\mu}}$. Therefore, $\{K_{\mu}=\sum_k \bra{\phi_{\mu}}U_k\ket{0} P_{\pi_{\mu}(k)}C^{\mu}P_k\}_\mu$ becomes a complete set of Strictly Block Incoherent Kraus operators. $\square$                                

\section{Physically Block Incoherent Operation}

For Physically Block Incoherent Operations (PBIO), we follow almost the procedure of physically incoherent operation of standard resource theory of coherence with significant changes. Here, we need to consider block incoherent unitary operators. Unlike PIO, after adding an incoherent ancilla to the original system, we apply a joint block incoherent unitary instead of standard incoherent unitary operators. Before going into our result, we state the following lemma, which is useful in this regard.\\
$\textbf{\textit{Lemma 3:}}$ A CPTP map defines PIO if and only if it can be expressed as a convex combination of maps, each having Kraus operators $\{K_j\}$ of the form $$K_j=U_jP_j=\sum_x e^{i\theta_x}\ket{\pi_j(x)}\bra{x}P_j$$
where $U_j$ are incoherent unitary, and all the $P_j$ form an orthogonal and complete set of incoherent projectors on system A and $\pi_j$ are permutations \cite{18}.\\

Now, we are ready to observe the structure of Kraus operators of Physically Block Incoherent Operations.

\vspace*{12pt}
\noindent
{\bf Theorem~4:}  Let $\mathcal{T}$ be a Completely Positive Trace Preserving map on S. $\mathcal{T}$ is a Physically Block Incoherent Operation if and only if it can be represented by a convex combination of maps, each having Kraus operators $\{K_y^{y'}\}$ such that each $K_y^{y'}$ has the form, $K_y^{y'}=U_y^{y'}P_y^{y'}$, where $U_y^{y'}$  are block incoherent unitary operators and $\{P_y^{y'}\}$ is a complete projective valued measurement of possibly higher ranks on system A.  

\vspace*{12pt}
\noindent
{\bf Proof:} Firstly, we consider the ancilla system to be in the state $\ket{y'}\bra{y'}$ and joint block incoherent unitary operator as
$$U_{AB}=\sum_i P_{\pi(i)}^{AB} \text{ } C \text{ } P_i^{AB},$$
Now if we take a higher rank projector $P_i^{AB}=\sum_{(\alpha,\alpha')\in S_i} \ket{\alpha \alpha'}\bra{\alpha \alpha'}$ of the joint system, and the complex matrix, $C=\sum_{s,s',t,t'} C_{ss'tt'}\ket{ss'}\bra{tt'}$, then $U_{AB}$ becomes
 $$U_{AB}= \sum_i \sum_{\substack{s,s',t,t'\\(s,s')\in S_{\pi(i)}\\ (t,t')\in S_i}} C_{ss'tt'}\ket{ss'}\bra{tt'},$$
 where $\{S_k=(i,j)\}_k$ is some partition of basis indices of the joint system and $\pi$ is a permutation. For example if we consider the basis $\{\ket{00},\ket{01},\ket{10},\ket{11}\}$ for two dimensional joint system and $S_0$ and $S_1$ as two partition of basis indices then $S_0$ and $S_1$ can be chosen as $S_0=\{(0,0), (0,1)\}$ and  $S_1=\{(1,0), (1,1)\}.$\\
\begin{equation*}
\begin{aligned}
&\;\;\text{Now, }U_{AB}(\rho_A \otimes \ket{y'}\bra{y'}) U_{AB}\\
&=\left(\sum_i \sum_{\substack{s,s',t,t'\\(s,s')\in S_{\pi(i)}\\ (t,t')\in S_i}} C_{ss'tt'}\ket{ss'}\bra{tt'}\right)(\rho_A \otimes \ket{y'}\bra{y'})\left(\sum_{i'} \sum_{\substack{s_1,s_1',t_1,t_1'\\(s_1,s_1')\in S_{\pi(i')}\\ (t_1,t_1')\in S_{i'}}} C_{s_1s_1't_1t_1'}\ket{s_1s_1'}\bra{t_1t_1'}\right)^\dagger\\
&=\sum_{i,i'}\sum_{\substack{s,s',t,t'\\(s,s')\in S_{\pi(i)}\\ (t,t')\in S_i}}  \sum_{\substack{s_1,s_1',t_1,t_1'\\(s_1,s_1')\in S_{\pi(i')}\\ (t_1,t_1')\in S_{i'}}} C_{ss'tt'}C_{s_1s_1't_1t_1'}^\dagger \bra{t}\rho_A \ket{t_1}\ket{s} \bra{s_1} \otimes \bra{t'} \ket{y'} \bra{y'}\ket{t_1'}\ket{s'}\bra{s_1'}\\
&=\sum_{i,i'} \sum_{\substack{s,s',t\\(s,s')\in S_{\pi(i)}\\ (t,y')\in S_i}}  \sum_{\substack{s_1,s_1',t_1\\(s_1,s_1')\in S_{\pi(i')}\\ (t_1,y')\in S_{i'}}} C_{ss'ty'}C_{s_1s_1't_1y'}^\dagger \bra{t}\rho_A \ket{t_1}\ket{s} \bra{s_1}\otimes \ket{s'}\bra{s_1'}
\end{aligned}
\end{equation*}

Next, we apply a rank one projective measurement in the incoherent basis $\{ \ket{y} \}$ to obtain the Kraus operator representation of the map. Now, the unnormalized state of the system becomes
\begin{equation*}
\begin{aligned}
&\bra{y}\left(U_{AB}(\rho_A \otimes \ket{y'}\bra{y'}) U_{AB}\right)\ket{y}\\ \\
&=\sum_{i,i'} \sum_{\substack{s,t\\(s,y)\in S_{\pi(i)}\\ (t,y')\in S_i}}  \sum_{\substack{s_1,t_1\\(s_1,y)\in S_{\pi(i')}\\ (t_1,y')\in S_{i'}}} C_{syty'}C_{s_1yt_1y'}^\dagger \bra{t}\rho_A \ket{t_1}\ket{s} \bra{s_1}\\ 
&=\left(\sum_{i} \sum_{\substack{s,t\\(s,y)\in S_{\pi(i)}\\ (t,y')\in S_i}}C_{syty'}\ket{s}\bra{t}\right)\rho_A\left(\sum_{i'}\sum_{\substack{s_1,t_1\\(s_1,y)\in S_{\pi(i')}\\ (t_1,y')\in S_{i'}}}C_{s_1yt_1y'}\ket{s_1}\bra{t_1}\right)^\dagger\\ \\
&=(K_y^{y'})\rho_A(K_y^{y'})^\dagger
\end{aligned}
\end{equation*}
$$\text{Here, each Kraus operator } K_y^{y'}=U_y^{y'}P_y^{y'} \text{ such that }
U_y^{y'}=\sum_{i} \sum_{\substack{s,t\\(s,y)\in S_{\pi(i)}\\ (t,y')\in S_i}}C_{syty'}\ket{s}\bra{t}+W_y^{y'},$$\\ where, the operator $W_y^{y'}$ is suitably chosen to make $U_y^{y'}$ a block unitary and $P_y^{y'}$ is a higher rank projective measurement,
$$P_y^{y'}=\sum_{i} \sum_{\substack{t\\(t,y)\in S_{\pi(i)}\\ (t,y')\in S_i}}\ket{t}\bra{t}.$$
Clearly, the set $\{K_y^{y'}\}_y$ is a complete set of  Kraus operators. If the ancilla system is chosen to be $\sum_{y'} p_{y'}\ket{y'}\bra{y'}$, then complete set of Kraus operators will be given as $\{ \sqrt{p_{y'}}K_y^{y'}\}_{y,y'}$ where each Kraus operator $K_{y}^{y'}$ has the previously described form.   $\square$

So, we get a complete characterization of Kraus operators of PBIO, which obviously has a block structure. Like PIO, each Kraus operator of PBIO is comprised of a block incoherent unitary and a higher dimensional projector. The only difference is that for PBIO, incoherent unitaries must be in block format, which clearly indicates that PBIO is a generalized version of PIO.
\section{Introduction to state transformation in block scenario}

The concept of state transformation under free operations is very important in every physically significant resource theory. Although many results of state transformation have been found under standard coherence theory \cite{26} till now, no result on state transformation has been established under the resource theory of block coherence. So, our main aim is to introduce concepts of state transformation under Strictly Block Incoherent Operation (SBIO).
We consider our input state as  $\ket{\psi}= \sum_i \psi_i \ket{i} $. Now for SBIO we take the CPTP map $\Lambda$ which has Kraus operator representation $\{ K_l \}_l$ such that $K_l= \sum_j P_{\pi_l(j)}C_l P_j$. Here, $P_j$ are projectors of higher rank, $C_l$ are arbitrary complex matrices, and $\pi_l$ are permutation functions.

According to our choice  $\{ K_l \}_l$ must satisfy the completeness condition, i.e., $\sum_l K_l^{\dagger} K_l=I$



$$\Rightarrow \sum_l \sum_j \sum_{\substack{i,k\in S_j\\ i' \in S_{\pi_l(j)}}}  \bra{i}C_l^{\dagger}\ket{i'} \bra{i'}C_l\ket{k} \ket{i}\bra{k}=I $$

So, for $i=k \in S_j$ (and hence j is fixed), we have,
\begin{equation}
\sum_l \sum_{i' \in S_{\pi_l(j)}} \mid \bra{i'}C_l\ket{i} \mid^2=1
\end{equation}
and for $i \neq k $,(but belong to same $S_j$ for some $j$),
\begin{equation}
\sum_l \sum_{i' \in S_{\pi_l(j)}} \bra{i}C_l^{\dagger}\ket{i'} \bra{i'}C_l\ket{k} =0
\end{equation}
For each $l$, we have 
\begin{equation}
    \begin{aligned}
        K_l\ket{\psi}&=\sum_j \left[ \left( \sum_{i\in S_{\pi_l(j)}}\ket{i}\bra{i}\right)C_l \left(  \sum_{i'\in S_{j}}\ket{i'}\bra{i'}\right)\right]\left( \sum_k \psi_k \ket{k}\right)\\
        &=\sum_j  \sum_{i\in S_{\pi_l(j)}} \sum_{k\in S_{j}} \psi_k \bra{i}C_l\ket{k} \ket {i}
    \end{aligned}
\end{equation}
\begin{equation*}
    \begin{aligned}
        \text{So, }\Lambda(\ket{\psi}\bra{\psi})&=\sum _l K_l\ket{\psi}\bra{\psi}K_l^\dagger\\
        &=\sum_l \sum_{j,j'} \sum_{\substack{i\in S_{\pi_l(j)}\\ i'\in S_{\pi_l(j')}}} \sum_{\substack{k\in S_j\\ k'\in S_{j'}}} \psi_k\psi_{k'}^* \bra{i}C_l\ket{k} \bra{i'}C_l\ket{k'}^*\ket{i}\bra{i'}
    \end{aligned}
\end{equation*}
where $C_l $ satisfies the above conditions (9) and (10).\\

Now we are particularly considering the case where the output state is a pure state $\ket{\phi}\bra{\phi}$, i.e., here we are focusing on the pure state transformation from $\ket{\psi}= \sum_i \psi_i \ket{i} $ to $\ket{\phi}= \sum_i \phi_i \ket{i} $ under the action of $\Lambda$.

So, $\Lambda(\ket{\psi}\bra{\psi})=\sum_{i,i'}\phi_i \phi_{i'}^*\ket{i}\bra{i'}$.

Comparing the coefficients, for fixed $i \in S_{\pi_l(j)}$ and $i'\in S_{\pi_l(j')}$,
$$\phi_i \phi_{i'}^*=\sum_{k,k' \in S_j} \psi_k \psi_{k'}^* \bra{i}C_l\ket{k}\bra{i'}C_l\ket{k'}^*$$
Then for $i=i'\in S_{\pi_l(j)}, \text{ we have }|\phi_i|^2=\sum_{k,k' \in S_j} \psi_k \psi_{k'}^* \bra{i}C_l\ket{k}\bra{i}C_l\ket{k'}^*$

So, coefficients of the output state can be calculated in the case of pure state transformation under SBIO. Again, we can further investigate this state transformation problem in an alternative way. If both $\ket{\psi}$ and $\ket{\phi}$ are fixed, then $\ket{\psi}$ can be transformed to $\ket{\phi}$ under SBIO if and only if there exists a solution for the above set of nonlinear equations together with the completeness conditions. Depending on the existence of the solution, we can conclude whether or not transformation between any two states is possible under SBIO.

Now, as an example, we discuss the above result elaborately for three-dimensional case by computing $\phi_0$, $\phi_1$ and $\phi_2$ explicitly. For the three dimensional case we consider two projectors $P_0$ and $P_1$ as $P_0= \ket{0}\bra{0}+\ket{1}\bra{1}$ and $P_1= \ket{2}\bra{2}$. We also choose permutation function as $\pi_0(0)=0$, $\pi_0(1)=1$, $\pi_1(0)=1$ and $\pi_1(1)=0$. Under these choices we can compute $\phi_0$, $\phi_1$ and $\phi_2$ as,

$\mid\phi_0\mid^2=(\psi_0\bra{0}C_0\ket{0}+\psi_1\bra{0}C_0\ket{1})^2+(\psi_2\bra{0}C_1\ket{2})^2$

$\mid\phi_1\mid^2=(\psi_0\bra{1}C_0\ket{0}+\psi_1\bra{1}C_0\ket{1})^2+(\psi_2\bra{1}C_1\ket{2})^2$

$\mid\phi_2\mid^2=(\psi_2\bra{2}C_0\ket{2})^2+(\psi_0\bra{2}C_1\ket{0}+\psi_1\bra{2}C_1\ket{1})^2$

Therefore, we achieve our goal of state transformation through free operations. This fact leads towards a new area of investigation. The next section will discuss different physical aspects of block coherence theory.

\section{Physical evidence enriching the structure of coherence theory}

In this section, our motivation is to connect the newly built concepts of several block incoherent operations with their operational significance and find some well-known applications in resource theory. It is well understood that the generalization of any resource theory is directly connected to the Naimark extension. Naturally, we have extended several free operations (IO, SIO, PIO) of coherence theory to their block representation (BIO, SBIO, PBIO) with proper justification. In this paper, a clear picture of Kraus operators has been provided precisely to represent a larger class of operations (BIO, SBIO, PBIO). During the development of coherence theory, it is noticeable that some physical properties of standard coherence theory may or may not be reflected in the generalized version. In the previous section, we explicitly establish the state transformation through SBIO. In future, this concept of state transformation may be generalized for other free operations.

Several important properties of standard coherence theory, like monogamy, distillation and distribution of coherence, can be reproduced in the light of block concepts. Different coherence measures can be redefined through the flavour of generalized coherence theory. Many concepts, protocols, and applications may be generalized in this similar passion. Through this process, it is expected that some new hidden physical aspects will evolve, which will further strengthen the development of the coherence theory.  
 
\section{Conclusion}
In this work, we have focused on the generalized structure of the resource theory of block coherence. We have introduced three free operations, namely, Block Incoherent Operation (BIO), Strictly Block Incoherent Operation (SBIO) and Physically Block Incoherent Operation (PBIO) and found analytical expressions for Kraus operators for them. We have proved that Kraus operators representing BIO have at most one non-zero block in each column partition, whereas Kraus operators representing SBIO have at most one non-zero block in each row and column partition. If we choose a block dephasing map with projectors of rank one, then our result matches the analytical representation of Kraus operators representing IO and SIO. To make this block coherence theory more physically motivated, we have included a dilation theorem, which can always be represented by Kraus operators representing SBIO. The characteristics of the Kraus operators representing PBIO have been elaborately discussed. Using the Naimark extension, we can define these free operations with respect to any POVM. To study the resource theory of POVM-based coherence, we need to consider the block formalism approach of Kraus operators representing various free operations in the extended Hilbert space. So, from this point of view, our work will be helpful. This will certainly help us to characterize free operations in this framework. In the future, one can further check whether other free operations like Dephasing-covariant Incoherent Operations(DIO) and Translationally Incoherent Operations (TIO) can be defined in block formalism. Finally, we discuss the state transformation protocol under SBIO as a physical application of the generalized coherence theory. We particularly calculated the output state generated from an input pure state under SBIO. Our work reveals that coefficients of the output state can be calculated in the case of pure state transformation. Again, in an alternative way, we present a set of nonlinear equations that must be satisfied for a pure state transformation under SBIO.  
Pure state transformation under SIO follows majorization criteria in standard resource theory. So, if we take the coefficients of both input and output pure states in our case to be real, then this will lead to further investigations of whether pure state transformations under SBIO satisfy majorization criteria or not. This clearly indicates a new area of research for the demand of a better understanding of coherence resource theory. 
\\
$\textbf{Note:}$ Recently, when we completed our work, we found a related work in \cite{27} where they have defined Block Incoherent Operations, Strictly Block Incoherent Operations and presented their Kraus operator forms with an analogy from standard coherence theory without proof. In our work, we have provided a full description.
\section*{ACKNOWLEDGEMENTS}
Priyabrata Char acknowledges the support from DST (Inspire), India. Also, the authors I. Chattopadhyay and D. Sarkar acknowledge the work as part of QUest initiatives by DST India.

\section*{Appendix-A}
\textbf{\large{Kraus operator formulation for Block Incoherent Operation for five-dimensional case:}} \\ 
We follow the same procedure as discussed in the main article. 
For a $5$ dimensional state $\rho$, a particular choice of block dephasing map can be made as\\
 $\Delta_1(\rho)=P_0 \rho P_0+P_1 \rho P_1$, where $P_0$ and $P_1$ are rank $2$ and rank $3$ projectors
 
For this particular choice of block dephasing map, equation (2) holds for two off-diagonal blocks of order $ 2 \times 3$ and $3 \times 2$, respectively and for $l=0$ and $l=1$, $d_{xy}^{l}$ are entries of diagonal blocks of order $ 2 \times 2$ and $3 \times 3$ respectively of a five-dimensional state.

\textbf{Case-I:} We consider the off-diagonal block of order $2 \times 3$. In this case, equation (2) holds for $i=0$ to $1$ and $i'=2$ to $4$. So, for the case of $\Delta_1$, we must have 
  \begin{equation}
  \sum_{x,y=0}^{1} c_{ix}^n d_{xy}^{0} c_{i^{'}y}^{n^*}=0
  \end{equation} and
  \begin{equation}
  \sum_{x,y=2}^{4} c_{ix}^n d_{xy}^{1} c_{i^{'}y}^{n^*}=0
  \end{equation} \\ \\ (9) gives the following sets of equations as 
  $$ c_{00}^n d_{00}^0 c_{20}^{n^*}+c_{01}^nd_{10}^0c_{20}^{n^*}+c_{00}^nd_{01}^0c_{21}^{n^*}+c_{01}^nd_{11}^0c_{21}^{n^*}=0$$
  $$ c_{00}^nd_{00}^0c_{30}^{n^*}+c_{01}^nd_{10}^0c_{30}^{n^*}+c_{00}^nd_{01}^0c_{31}^{n^*}+c_{01}^nd_{11}^0c_{31}^{n^*}=0$$
  $$c_{00}^nd_{00}^0c_{40}^{n^*}+c_{01}^nd_{10}^0c_{40}^{n^*}+c_{00}^nd_{01}^0c_{41}^{n^*}+c_{01}^nd_{11}^0c_{41}^{n^*}=0$$
  $$c_{10}^nd_{00}^0c_{20}^{n^*}+c_{11}^nd_{10}^0c_{20}^{n^*}+c_{10}^nd_{01}^0c_{21}^{n^*}+c_{11}^nd_{11}^0c_{21}^{n^*}=0$$
  $$c_{10}^nd_{00}^0c_{30}^{n^*}+c_{11}^nd_{10}^0c_{30}^{n^*}+c_{10}^nd_{01}^0c_{31}^{n^*}+c_{11}^nd_{11}^0c_{31}^{n^*}=0$$
   $$c_{10}^nd_{00}^0c_{40}^{n^*}+c_{11}^nd_{10}^0c_{40}^{n^*}+c_{10}^nd_{01}^0c_{41}^{n^*}+c_{11}^nd_{11}^0c_{41}^{n^*}=0$$ 
  and (10) gives the following set of equations 
 \begin{equation*}
    \begin{aligned}
c_{02}^nd_{22}^1c_{22}^{n^*}+c_{02}^nd_{23}^1c_{23}^{n^*}+c_{02}^nd_{24}^1c_{24}^{n^*}+c_{03}^nd_{32}^1c_{22}^{n^*}+c_{03}^nd_{33}^1c_{23}^{n^*}&+c_{03}^nd_{34}^1c_{24}^{n^*}+c_{04}^nd_{42}^1c_{22}^{n^*}\\&+c_{04}^nd_{43}^1c_{23}^{n^*}+c_{04}^nd_{44}^1c_{24}^{n^*}=0 
     \end{aligned}
\end{equation*}
 \begin{equation*}
    \begin{aligned}
c_{02}^nd_{22}^1c_{32}^{n^*}+c_{02}^nd_{23}^1c_{33}^{n^*}+c_{02}^nd_{24}^1c_{34}^{n^*}+c_{03}^nd_{32}^1c_{32}^{n^*}+c_{03}^nd_{33}^1c_{33}^{n^*}&+c_{03}^nd_{34}^1c_{34}^{n^*}+c_{04}^nd_{42}^1c_{32}^{n^*}\\&+c_{04}^nd_{43}^1c_{33}^{n^*}+c_{04}^nd_{44}^1c_{34}^{n^*}=0 
 \end{aligned}
\end{equation*}
\begin{equation*}
    \begin{aligned}
c_{02}^nd_{22}^1c_{42}^{n^*}+c_{02}^nd_{23}^1c_{43}^{n^*}+c_{02}^nd_{24}^1c_{44}^{n^*}+c_{03}^nd_{32}^1c_{42}^{n^*}+c_{03}^nd_{33}^1c_{43}^{n^*}&+c_{03}^nd_{34}^1c_{44}^{n^*}+c_{04}^nd_{42}^1c_{42}^{n^*}\\&+c_{04}^nd_{43}^1c_{43}^{n^*}+c_{04}^nd_{44}^1c_{44}^{n^*}=0 
 \end{aligned}
\end{equation*}
 \begin{equation*}
    \begin{aligned}
c_{12}^nd_{22}^1c_{22}^{n^*}+c_{12}^nd_{23}^1c_{23}^{n^*}+c_{12}^nd_{24}^1c_{24}^{n^*}+c_{13}^nd_{32}^1c_{22}^{n^*}+c_{13}^nd_{33}^1c_{23}^{n^*}&+c_{13}^nd_{34}^1c_{24}^{n^*}+c_{14}^nd_{42}^1c_{22}^{n^*}\\&+c_{14}^nd_{43}^1c_{23}^{n^*}+c_{14}^nd_{44}^1c_{24}^{n^*}=0 
 \end{aligned}
\end{equation*}
\begin{equation*}
    \begin{aligned}
c_{12}^nd_{22}^1c_{32}^{n^*}+c_{12}^nd_{23}^1c_{33}^{n^*}+c_{12}^nd_{24}^1c_{34}^{n^*}+c_{13}^nd_{32}^1c_{32}^{n^*}+c_{13}^nd_{33}^1c_{33}^{n^*}&+c_{13}^nd_{34}^1c_{34}^{n^*}+c_{14}^nd_{42}^1c_{32}^{n^*}\\&+c_{14}^nd_{43}^1c_{33}^{n^*}+c_{14}^nd_{44}^1c_{34}^{n^*}=0 
 \end{aligned}
\end{equation*}
  \begin{equation*}
\begin{aligned}
c_{12}^nd_{22}^1c_{42}^{n^*}+c_{12}^nd_{23}^1c_{43}^{n^*}+c_{12}^nd_{24}^1c_{44}^{n^*}+c_{13}^nd_{32}^1c_{42}^{n^*}+c_{13}^nd_{33}^1c_{43}^{n^*}&+c_{13}^nd_{34}^1c_{44}^{n^*}+c_{14}^nd_{42}^1c_{42}^{n^*}\\&+c_{14}^nd_{43}^1c_{43}^{n^*}+c_{14}^nd_{44}^1c_{44}^{n^*}=0
  \end{aligned}
\end{equation*}
  
To find entries $d_{xy}^0$ and $d_{xy}^1$, the particular choice of arbitrary states can be made as $\frac{\ket{0}+\ket{1}}{\sqrt{2}}$ , $\frac{\ket{0}+\ket{2}}{\sqrt{2}}$, $\frac{\ket{0}+\ket{3}}{\sqrt{2}}$, $\frac{\ket{0}+\ket{4}}{\sqrt{2}}$, $\frac{\ket{1}+\ket{2}}{\sqrt{2}}$ , $\frac{\ket{2}+\ket{3}}{\sqrt{2}}$, $\frac{\ket{2}+\ket{4}}{\sqrt{2}}$, $\frac{\ket{3}+\ket{4}}{\sqrt{2}}$. Now we decompose these states into two blocks by applying $\Delta_1$, where entries of block of order $ 2 \times 2$ give values of $d_{xy}^0$ and entries of block of order $3 \times 3$ give values of $d_{xy}^1$. After that, we substitute $d_{xy}^{0}$ and $d_{xy}^{1}$ in the sets of equations (10) and (11). After solving, from (10) we can show that if we consider any one of $c_{00}^n$, $c_{01}^n$, $c_{10}^n$, $c_{11}^n$ to be non-zero then $c_{20}^n=c_{30}^n=c_{40}^n=c_{21}^n=c_{31}^n=c_{41}^n=0$ and vice versa. From (11) we can similarly show that if any one of $c_{02}^n$, $c_{03}^n$, $c_{04}^n$, $c_{12}^n$, $c_{13}^n$, $c_{14}^n$ to be non-zero then $c_{22}^n=c_{23}^n=c_{24}^n=c_{32}^n=c_{33}^n=c_{34}^n=c_{42}^n=c_{43}^n=c_{44}^n=0$ and vice versa. 

Thus, for each n, Kraus operator $K_n$ has any one  of the following form\\
 $$K_n=
\left[
\begin{array}{c|c}
* & * \\
\hline
\bf 0 & \bf 0
\end{array}
\right]
,\quad 
K_n=
\left[
\begin{array}{c|c}
\bf 0 & \bf 0 \\
\hline
* & *
\end{array}
\right]
, \quad
K_n=
\left[
\begin{array}{c|c}
\bf 0 & * \\
\hline
* & \bf 0
\end{array}
\right]
, \quad
{{K_n}}=
\left[
\begin{array}{c|c}
* & \bf 0 \\
\hline
\bf 0 & *
\end{array}
\right]
.$$
where * indicates that the block may have some non-zero entry and \textbf{0} is the block with zero entries.

\textbf{Case-II:} Now we are considering the case of an off-diagonal block of order $ 3 \times 2$. In this case, we can get the same result if we proceed in the same way as discussed in Case I.\\

\section*{Appendix-B} 
\textbf{\large{Kraus operator formulation for Strictly Block Incoherent Operation for five-dimensional case:}} \\ \\   
   For the choice of block dephasing map $\Delta_1$, equation (5) holds for two diagonal blocks of order $ 2 \times 2$ and $3 \times 3$ and the entries $f_{xy}^{01}$ and $f_{xy}^{10}$ can be found by applying $P_0 \sigma P_1$ and $P_1 \sigma P_0$ respectively, where $P_0$ and $P_1$ are projectors of rank 2 and 3, $\sigma$ is arbitrary five-dimensional state.
\\
  Thus for the choice of $\Delta_1$ we have,
   \begin{equation} 
   \sum_{x=0}^{1} \sum_{y=2}^{4} c_{ix}^n f_{xy}^{01} c_{i'y}^{n^*}=0
   \end{equation}
   \begin{equation}
   \sum_{x=2}^{4} \sum_{y=0}^{1} c_{ix}^n f_{xy}^{10} c_{i'y}^{n^*}=0
   \end{equation} for $i, i'=0$ to $1$ and
   \begin{equation} 
   \sum_{x=0}^{1} \sum_{y=2}^{4} c_{ix}^n f_{xy}^{01} c_{i'y}^{n^*}=0
   \end{equation}
   \begin{equation}
   \sum_{x=2}^{4} \sum_{y=0}^{1}  c_{ix}^n f_{xy}^{10} c_{i'y}^{n^*}=0
   \end{equation}for $i, i'=2$ to $4$. \\\\
   Proceeding in the way just like the case of Block Incoherent Operation, we can make choice of $\sigma$ as $\frac{\ket{0}+\ket{1}}{\sqrt{2}}$, $\frac{\ket{0}+\ket{2}}{\sqrt{2}}$, $\frac{\ket{0}+\ket{3}}{\sqrt{2}}$, $\frac{\ket{0}+\ket{4}}{\sqrt{2}}$, $\frac{\ket{1}+\ket{2}}{\sqrt{2}}$, $\frac{\ket{1}+\ket{3}}{\sqrt{2}}$, $\frac{\ket{1}+\ket{4}}{\sqrt{2}}$.  After taking two off-diagonal blocks of given states, we substitute values of $f_{xy}^{01}$ and $f_{xy}^{10}$ in the set of equations (11), (13) and (12),(14) respectively. Solving this from (11), we can show that if we choose any one of $c_{00}^n$, $c_{01}^n$, $c_{10}^n$, $c_{11}^n$ in non-zero then $c_{02}^n=c_{03}^n=c_{04}^n=c_{12}^n=c_{13}^n=c_{14}^n=0$ and vice versa. Similarly from (13) we can show that if any one of $c_{20}^n$, $c_{21}^n$, $c_{30}^n$, $c_{31}^n$, $c_{40}^n$, $c_{41}^n$ is non-zero then $c_{22}^n=c_{23}^n=c_{24}^n=c_{32}^n=c_{33}^n=c_{34}^n=c_{42}^n=c_{43}^n=c_{44}^n=0$ and vice versa. We can get the same results if we solve the set of equations (12) and (14). 
 In this way, we have shown that every Kraus operator representing Strictly Block Incoherent Operation has at most one non-zero block in each row partition. Since a Strictly Block Incoherent Operation is also a Block Incoherent Operation, it must have at most one non-zero block in each column partition.
 Thus, for each n, Kraus operator $K_n$ has any one of the following form 
 $$K_n=
\left[
\begin{array}{c|c}
* & 0 \\
\hline
0 & *
\end{array}
\right]
, K_n=
\left[
\begin{array}{c|c}
0 & * \\
\hline
* & 0
\end{array}
\right].
$$

\end{document}